\documentclass[a4paper,11pt]{article}
\pdfoutput=1 % if your are submitting a pdflatex (i.e. if you have
\usepackage{jheppub} % for details on the use of the package, please
\usepackage[T1]{fontenc} % if needed

\usepackage{pdfsync}
\usepackage{mathptmx}
\usepackage[utf8]{inputenc}
\usepackage[T1]{fontenc}
\usepackage{slashed}
\usepackage{times}
\usepackage{amssymb,amsfonts,amsmath,amsthm}
\usepackage{dsfont,bbm}
\usepackage{dcolumn}
\usepackage{epsf}
\usepackage{graphicx}
\usepackage[caption=false]{subfig}
\usepackage{dsfont}
\usepackage{bm}
\usepackage{tikz}
\usepackage[matrix, arrow, curve]{xy}
\usepackage{slashed}
\usepackage[active]{srcltx}

\newcommand{\up}{{\uparrow}}

\title{\boldmath On new $k_\perp$-dependent (quasi)parton distribution functions}

\author[a]{I.~V.~Anikin}

\affiliation[a]{Bogoliubov Laboratory of Theoretical Physics, JINR, 141980 Dubna, Russia}

\emailAdd{anikin@theor.jinr.ru}

\abstract{
We advocate the existence  of a new type of $k_\perp$-dependent functions.
In contrast to the well-known Boer-Mulders function,
the presented new functions can be associated with the collective alignment of quark spin vectors.
Moreover, the new functions are sensitive to the transverse motion of partons inside hadrons, which
are linked to the spin alignment of partons, 
and they are initiated by the interactions encoded in the corresponding correlators.

\vspace{10cm}
\noindent
{\footnotesize Based on the materials presented at the “International Conference on Quantum Field Theory, High-Energy Physics, and Cosmology”
held at the Bogoliubov Laboratory of Theoretical Physics of the JINR (Dubna, Russian Federation), July 18th - 21st, 2022.}
}

\begin{document}
\maketitle
\flushbottom

\vspace{0.5cm}
%%%%%%%%%%%%%%%%%%%%%%%%%%%%%%%%%%%%
\section{Instead of Introduction}
\label{Intro}
%%%%%%%%%%%%%%%%%%%%%%%%%%%%%%%%%%%%

Nowadays, it is very well understood that both the inclusive and exclusive reactions
play an important role in getting of substantial information on the inner composite structure
of hadrons. In particular, it refers to the reactions under the special asymptotical regime
where the corresponding transverse momentum is (very) large.
Such a kind of processes has being called as the hard reactions.
Traditionally, the large transverse momentum is provided by the large (off-shellness) virtuality
of photons, see for example the Drell-Yan-like and/or Compton-like processes.
The attractive point of any hard reactions is that the given processes can be investigated within the
factorization procedure. The mathematical basis of every factorization has been formed with the help
of the factorization theorem that is proven (or has to be proven) in QCD (or in the standard model).
Symbolically, it can be presented as the following
\begin{eqnarray}
\hspace{-1cm}
\xymatrix{
&\boxed{Exclusive \,\, and\,\, Inclusive \,\,Reactions} \ar@{=>}[d]^{Q^2\to\infty}&\\
&\boxed{Hard \,\,Reactions.}\ar@{=>}[d]^{Factorization\,\,Theorem}_{(asympt.\,\, estimations)}&\\
&\boxed{\Big[E(x_1, ....) \otimes \Phi(x_1, ....)\Big]
\oplus
\Big[\Phi(y_1, ....) \otimes E(x_1, y_1; ....) \otimes \Phi(x_1, ....)\Big] }}
\nonumber
\end{eqnarray}
Here, $E(....)$ and $\Phi(....)$
denote the hard (perturbative) and soft (non-perturbative) parts, respectively.

One can see that the amplitude (or hadron tensor) representation in the form of
the mathematical convolution $E(x_1, ....) \otimes \Phi(x_1, ....)$ is entirely a result
of the corresponding asymptotical estimation. In other words, the asymptotical estimation of the
given amplitudes/hadron tensors, instead of the direct calculations of amplitudes, leads to
the factorized convolution (or to factorization) of the physically independent functions $E(x_1, ....)$
and $\Phi(x_1, ....)$.

If the structure of the hard part function, which is  $E(x_1, ....)$, is fixed by the perturbative calculations,
the soft part function given by $\Phi(x_1, ....)$ is pure non-perturbative and is
not calculable in the standard model unless the effective Lagrangian of quark-hadron interactions has been introduced
within the effective model. On the other hand, the properties of  $\Phi(x_1, ....)$ and its different representations which stem
from the fundamental symmetries and Lorentz covariance are under the intensive studies.
The different functions Lorentz-parametrizing the soft part function $\Phi(x_1, ....)$ can be associated with
the probability amplitudes (or their extensions) which describe the different distributions of partons inside hadrons.
The parton distribution interpretation gives the connection with the experimental data in a sense that
the experimental data analysis include the fitting parameters (for example, the correlators of local Gegenbauer operators)
which can be traced from the parton distributions
(see, for example,
\cite{Boglione:1999pz, Bacchetta:2004jz, Goeke:2005hb, Collins:2005rq, Anselmino:2008sga, Bastami:2018xqd,
Boer:1997nt}).

In the simplest {\it (a)} case of collinear and non-interacting partons, {\it i.e.} the dependence of $k_\perp$ is absent ($k_\perp=0$)
and $\mathbb{S}$-matrix is trivial ($\mathbb{S}=\mathbb{I}$) in the corresponding correlators,
the parton distributions have the well-defined probability interpretations in terms of the corresponding probability amplitudes.
In this case, the parton distributions provide the ``statical'' probability description, {\it i.e.} without any evolutions.
If we deal with the small $k_\perp$-dependence in the correlators, $|k_\perp|= {\cal O}(k^2_\perp)$, and
the interactions have been involved in the consideration ($\mathbb{S}\not=\mathbb{I}$),
the evolution equations of parton distributions can be available for the investigations. This is the case {\it (b)}.
Here, we still have the probability interpretations
of parton distributions.
However, in the {\it (c)} case of substantial $k_\perp$-dependence of the paramtrizing functions (or in the corresponding correlators),
{\it i.e.} $k_\perp\not=0$ together with the non-trivial interactions $\mathbb{S}\not=\mathbb{I}$, the probability interpretation of
$k_\perp$-dependent parton distributions does not take place directly, but it can be restored after the $k_\perp$-integration
(with the corresponding weight functions), {\it i.e.} for the function moments.
Besides, this non-trivial case can lead to the existence of a new type of $k_\perp$-dependent parametrizing functions
due to the interactions which are presented in the correlators.

Indeed, as demonstrated below, the Lorentz parametrization of a given correlator has being formed by the
set of external vectors which are describing the hadron and the set of internal vectors
which are describing the partons inside hadrons.
If the external vectors are fixed, the set of internal vectors depends
on the interactions (or the order of interactions) which are encoded in the correlator.
It is clear that the interactions have to be presented in the correlators, otherwise there is no a possibility, at least,
to calculate the parton evolution.
At the same time, the interactions at the given order can be spit onto
the evolution type (this type does not give new functions) and the structure type (this type gives new functions).

Schematically, we can represent three mentioned cases as
 \begin{eqnarray}
\hspace{-2cm}
\xymatrix{
\hspace{1.3cm}&&
\boxed{\Big[E(....) \otimes \Phi(....)\Big]}
&&\\
\hspace{1.3cm}\boxed{ PDs\,\, (no \,\,EE)}
\ar@/^1pc/@{<=}[ur]^{k_\perp=0}_{\mathbb{S}=\mathbb{I}}
% \ar@/_1pc/@{<=}[ur]^{k_\perp=0}_{\mathbb{S}=\mathbb{I}}
&&
\boxed{PDs \,\,(EE)}   \ar@{<=}[u]^{k_\perp\approx 0}_{\mathbb{S}\not=\mathbb{I}}
&&
\boxed{new \,\, PDs \,\,(EE)}
\ar@/_1pc/@{<=}[ul]^{ k_\perp\not=0}_{ \mathbb{S}\not=\mathbb{I}}
%\ar@{<=}[ul]^{k_\perp\not=0}_{\mathbb{S}\not=\mathbb{I}}
}
\nonumber
\end{eqnarray}

In all three cases, the hadron momentum $P$ and the hadron spin $S$ are the external vectors the number of which are usually unchanged.
Meanwhile, the case {\it (c)} provides the extended set of internal vectors, $\{ k^+, s^+, k_\perp, s_\perp \}$, in contrast to
the cases {\it (a)} and {\it (b)} where the internal vectors are $\{ k^+, s^+ \}$ only. Here, the plus denotes the plus light-cone component
of the given vector. By definition, the internal vectors (as the internal parameters) are related to the parton operators which form the
given correlators. For example, the correlator given by
\begin{eqnarray}
\label{Intro-1}
\langle P, S | \bar\psi \gamma^\mu (1+\gamma_5) \psi | P, S \rangle
\end{eqnarray}
involves the set of internal vectors is defined by the operator $\bar\psi \gamma^\mu (1+\gamma_5) \psi$.
Namely, one writes the following
(the corresponding Fourier transforms denoted by $\stackrel{{\cal F}}{\Longrightarrow}$
for the fermion field operators)
\begin{eqnarray}
\label{Intro-2}
&&- \, \bar\psi^{(s)}\gamma^\mu (1+\gamma_5) \psi^{(s)} =
\text{tr} \big[ \psi^{(s)} \bar\psi^{(s)} \gamma^\mu (1+\gamma_5) \big]
\stackrel{{\cal F}}{\Longrightarrow}
\nonumber\\
&&
\frac{1}{4}\text{tr} \big[ (\hat k + m_q) (1+ \gamma_5 \lambda_s +
\gamma_5 \hat s) \gamma^\mu (1+\gamma_5) \big]
= k^\mu - \lambda_s k^\mu + m_q s^\mu.
\end{eqnarray}
From that, we can see that the parton operator of Eqn.~(\ref{Intro-1}) gives two internal vectors $\{ k, s\}$.
So, the vectors $\{ P, S\}$ and  $\{ k, s\}$ can participate in the Lorentz parametrization.
In the case of correlators with the included interactions, the situation is similar but more tricky.

%%%%%%%%%%%%%%%%%%%%%%%%%%%%%%%%%%%%
\section{The role of interactions in the correlators}
\label{Inter}
%%%%%%%%%%%%%%%%%%%%%%%%%%%%%%%%%%%%

In this section we demonstrate the role of interaction in the corresponding correlator \cite{Anikin:2021zxl, Anikin:2022ocg, Anikin:2022eyf}.
For the pedagogical reason, we begin with the forward Compton scattering (CS) amplitude which reads
\begin{eqnarray}
\label{Amp-1}
{\cal A}_{\mu\nu} = \langle P| a^-_\nu(q) \, \mathbb{S}[\bar\psi, \psi, A] \, a^+_\mu(q) | P\rangle,
%\nonumber
\end{eqnarray}
where $\mathbb{S}$-matrix defined as
\begin{eqnarray}
\mathbb{S}[\psi,\bar\psi, A]={\rm T}\, \text{exp}\Big\{
i \int (d^4 z) \big[ {\cal L}_{QCD}(z) + {\cal L}_{QED}(z)\big] \Big\}.
%\nonumber
\end{eqnarray}
Making used the commutation relations of creation (or annihilation) operators with $\mathbb{S}$-matrix
\begin{eqnarray}
\label{Com-1}
&&\big[ a^\pm_\mu(q), \, \mathbb{S}[\bar\psi, \psi, A] \big]= \int (d^4 \xi) e^{\pm i q \xi}
\frac{\delta \mathbb{S}[\bar\psi, \psi, A] }{ \delta A^\mu(\xi)}, \quad \text{where}
\nonumber\\
&&
\label{Der-S}
\frac{\delta \mathbb{S}[\bar\psi, \psi, A] }{ \delta A^\mu(\xi)} =
{\rm T} \, \Big\{  \int (d^4 z) \frac{\delta {\cal L}_{QED}(z)}{\delta A^\mu(\xi)} \, \mathbb{S}[\bar\psi, \psi, A] \Big\},
%\nonumber
\end{eqnarray}
the CS-amplitude can be rewritten as
\begin{eqnarray}
\label{Amp-2}
&&{\cal A}_{\mu\nu} =
\int (d^4 \xi_1) (d^4 \xi_2) e^{- i q (\xi_1 - \xi_2)}
\langle P| \frac{\delta^2 \mathbb{S}[\bar\psi, \psi, A] }{ \delta A^\mu(\xi_1) \delta A^\nu(\xi_2)} | P\rangle
\nonumber\\
&&
\Rightarrow \int (d^4 z) e^{- i q z} \langle P| \, {\rm T}
\Big\{ [\bar\psi(0)\gamma_\nu \psi(0)] \,
[\bar\psi(z)\gamma_\mu \psi(z)]\,
\mathbb{S}[\bar\psi, \psi, A]  \Big\}
| P\rangle.
%\nonumber
\end{eqnarray}
Using Wick's theorem and calculating the only quark operator contraction,
we get the simplest ``hand-bag'' diagram contribution to the CS-amplitude.
It reads
\footnote{$\delta^{(4)} (\text{momentum conserv.})$, as a common prefactor, is not shown.}
\begin{eqnarray}
\label{Amp-3}
&&{\cal M}_{\mu\nu}^{\text{hand-bag}} =
\int (d^4 k) \,\text{tr} \big[ E_{\mu\nu}(k) \Phi(k) \big],
%\nonumber
\end{eqnarray}
where
\footnote{Here, the subscript $``c"$ denotes the connected diagram contributions which we only consider.}
\begin{eqnarray}
\label{E}
&&E_{\mu\nu}(k) = \gamma_\mu S(k+q) \gamma_\nu + \gamma_\nu S(k-q) \gamma_\mu,
\nonumber\\
&&
\label{phifun}
\Phi(k)= \int (d^4 z) \, e^{ikz}
\langle P| \,\widetilde{{\rm T}} \, \bar\psi(0) \psi(z) \mathbb{S}[\bar\psi, \psi, A] | P \rangle_c.
%\nonumber
\end{eqnarray}
In order to compactify the representation of $\Phi(k)$, we can go over to the Heisenberg representation of correlators, {\it i.e.}
\begin{eqnarray}
\label{phifun-H}
\Phi(k)= \int (d^4 z) \, e^{ikz}
\langle P| \bar\psi(0) \psi(z) | P \rangle^H.
%\nonumber
\end{eqnarray}
Notice that the factorization procedure is not yet applied to the CS-amplitude, see Eqn.~(\ref{Amp-3}).

Since the amplitude ${\cal M}_{\mu\nu}$ involves the non-perturbative correlator, $\langle P | {\cal O}(0,z) | P\rangle$,
it cannot be calculated within the perturbative (standard) theory of QCD (pQCD). Indeed, the hadron state is given by
$ | P \rangle = \mathbf{a}^+_h(\psi, \bar\psi | A) \, | 0 \rangle$
where the hadron operator  $\mathbf{a}^+_h(\psi, \bar\psi | A)$ is undefined in pQCD. As a result,
anticommutator $\big[ \, \psi(0), \, \mathbf{a}^+_h(\psi, \bar\psi | A)\, \big]_+$ remains unknown and
the direct calculation of amplitude is not available (unless we exceed the frame of pQCD introducing the effective
quark-hadron Lagrangian \cite{Efimov:1993zg}).
However,  instead of direct calculation,  ${\cal M}_{\mu\nu}$ can be estimated with the help of the suitable asymptotical regime,
$q^2=-Q^2 \to\infty$. The estimation of amplitudes within the asymptotical regime is a object of factorization procedure, see below.

%%%%%%%%%%%%%%%%%%%%%%%%%%%%%%%%%%%%
\section{Factorization theorem (factorization procedure)}
\label{FT}
%%%%%%%%%%%%%%%%%%%%%%%%%%%%%%%%%%%%

%%%%%%%%%%%%%%%%%%%%%%%%%%%%%%%%%%%%
\subsection{The forward CS-amplitude}
\label{FT:I}
%%%%%%%%%%%%%%%%%%%%%%%%%%%%%%%%%%%%

To outline the (typical) factorization procedure, we consider the CS-like amplitude written in the momentum representation, we have the
following (cf. Eqn.~(\ref{Amp-3}))
\footnote{Here, for the sake of brevity, we omit all possible Lorentz indices.}
\begin{eqnarray}
\label{Fact-1}
A=\int (d^4 k) E(k, q) \, \Phi(k),
%\nonumber
\end{eqnarray}
where $E(k, q) $ is given by the propagator product and
\begin{eqnarray}
\label{Phi-1q}
\Phi(k)\stackrel{{\cal F}}{=} \langle \bar\psi(z) \Gamma \psi (0) \rangle,
%\nonumber
\end{eqnarray}
with $\stackrel{{\cal F}}{=}$ denoting the Fourier transform.
In Eqn.~(\ref{Phi-1q}),
we do not specify the presence of interaction in the correlator for a while
because we want to avoid the extra complications for this discussion.

We now have to choose the dominant directions dictated by the given process kinematics.
For CS-amplitude, we deal with the only dominant directions
associated with the plus light-cone direction.
We then have to introduce the definitions of the dimensionless parton fractions as
\begin{eqnarray}
\label{Fth-Rep}
d^4 k \Rightarrow d^4 k \int_{-1}^{+1}dx \delta(x - k^{+}/P^{+})
%\nonumber
\end{eqnarray}
and to expand $E(k, q)$ around the chosen dominant direction.
As a result, we obtain that
\begin{eqnarray}
\label{Fact-2}
A^{(0)}=\int (d x) \, E(xP^+ ; q)
\Big\{
\int (d^4 k) \delta(x - k^{+}/P^{+})
\Phi(k)
\Big\}
\end{eqnarray}
 if  $k^\perp_i$-terms are neglected in the expansion;
 and
 \begin{eqnarray}
\label{Fact-3}
A^{(k_\perp)}=\int (d x) \sum_{i}E^{(i)}(xP^+ ; q)
\Big\{
\int (d^4 k) \delta(x - k^{+}/P^{+}) \,
\prod_{i^\prime=1}^{i}
k_{i^\prime}^\perp\,
\Phi(k)
\Big\}
\end{eqnarray}
if $k_\perp$-terms are essential in the expansion.

%%%%%%%%%%%%%%%%%%%%%%%%%%%%%%%%%%%%
\subsection{The Drell-Yan-like hadron tensor}
\label{FT:II}
%%%%%%%%%%%%%%%%%%%%%%%%%%%%%%%%%%%%

In the similar manner, we can treat the DY-like hadron tensor. Before factorization, it reads
\begin{eqnarray}
\label{Fact-1}
W=\int (d^4 k_1) (d^4 k_2) E(k_1, k_2, q) \, \Phi_1(k_1) \, \bar\Phi_2(k_2),
%\nonumber
\end{eqnarray}
where
\begin{eqnarray}
\label{E-Phi}
&&E(k_1, k_2, q) = \delta^{(4)}(k_1+k_2-q) \, {\cal E}(k_1, k_2, q)
\nonumber\\
&&
\Phi_1(k_1)\stackrel{{\cal F}_1}{=} \langle \bar\psi(z_1) \Gamma_1 \psi (0) \rangle,
\quad
\bar\Phi_2(k_2)\stackrel{{\cal F}_2}{=} \langle \bar\psi(0) \Gamma_2 \psi (z_2) \rangle
%\nonumber
\end{eqnarray}
and $\stackrel{{\cal F}_i}{=}$ denotes the corresponding Fourier transforms.
As a result of factorization, we obtain that
\begin{eqnarray}
\label{Fact-2}
&&W^{(0)}=\int (d x_1) (d x_2) E(x_1P^+_1, x_2P^-_2; q)
\Big\{
\int (d^4 k_1) \delta(x_1 - k_1^{+}/P_1^{+})
\Phi_1(k_1)
\Big\}
\nonumber\\
&&
\times
\Big\{
\int (d^4 k_2)  \delta(x_2 - k_2^{-}/P_2^{-})
\bar\Phi_2(k_2)
\Big\}
%\nonumber
\end{eqnarray}
for the unessential (integrated out in the soft functions) $k_\perp$-dependence;
 and
 \begin{eqnarray}
\label{Fact-3}
&&W^{(k_\perp)}=\int (d x_1) (d x_2) \sum_{i,j}E^{(i,j)}(x_1P^+_1, x_2P^-_2; q)
\Big\{
\int (d^4 k_1) \delta(x_1 - k_1^{+}/P_1^{+}) \,
\prod_{i^\prime=1}^{i}
k_{1\, i^\prime}^\perp\,
\Phi_1(k_1)
\Big\}
\nonumber\\
&&
\times
\Big\{
\int (d^4 k_2)  \delta(x_2 - k_2^{-}/P_2^{-})\,
\prod_{j^\prime=1}^{j}
k_{2\, j^\prime}^\perp\,
\bar\Phi_2(k_2)
\Big\}
%\nonumber
\end{eqnarray}
for the essential $k_\perp$-dependence in the soft functions.

It is worth to notice that in contrast to the genuine-factorized forms, we adhere in this paper,
the approaches with $q_\perp\not=0$
and without the $\delta$-function expansion result in
\begin{eqnarray}
\label{NoFact-1}
&&\tilde W^{(0)}=\int (d x_1) (d x_2) {\cal E}(x_1P^+_1, x_2P^-_2; q)
\Big\{
\int (d^2 \vec{\bf k}^\perp_1)
(d^2 \vec{\bf k}^\perp_2)
\delta^{(2)} (\vec{\bf k}^\perp_1 + \vec{\bf k}^\perp_2 - \vec{\bf q}^\perp)
\nonumber\\
&&
\times
\int (dk^+_1 d k_1^-) \delta(x_1 - k_1^{+}/P_1^{+})
\Phi_1(k_1)
\nonumber\\
&&
\times
\int (d k^-_2 dk^+_2)  \delta(x_2 - k_2^{-}/P_2^{-})
\bar\Phi_2(k_2)
\Big\},
%\nonumber
\end{eqnarray}
where $\Phi(k_1)$ and $\bar\Phi(k_2)$ cannot be independent each others due to the linking $k_\perp$-integrations.
The representation of Eqn.~(\ref{NoFact-1}) leads to the factorization breaking effects which should be compensated by, for example,
$e^{-S(\vec{\bf k}^2_\perp/\Lambda^2)}$-multiplication minimizing the non-factorized effects.

%%%%%%%%%%%%%%%%%%%%%%%%%%%%%%%%%%%%
\section{Infinite momentum frame: the collinear limit, $k_\perp=0$, and $\mathbb{S}=\mathbb{I}$}
\label{ColF}
%%%%%%%%%%%%%%%%%%%%%%%%%%%%%%%%%%%%

We now return to the discussion of CS-amplitude, see Eqn.~(\ref{Amp-3}). Let us assume that
$\mathbb{S}=\mathbb{I}$ (no QCD interactions) for the hand-bag diagram, we have the following
\begin{eqnarray}
\label{Amp-2-2}
{\cal A}_{\mu\nu}\Big|_{\mathbb{S}=\mathbb{I}} =
\int (d^4 z_1\, d^4 z_2) \, e^{- i q (z_1 - z_2)} \langle P| \,
 \textbf{:} \bar\psi(z_1) \, E_{\mu\nu}(z_1-z_2) \,\psi(z_2)  \textbf{:} \,
| P\rangle.
%\nonumber
\end{eqnarray}
Focusing on $\gamma^+$-projection in the correlator within the momentum representation, we get that
\begin{eqnarray}
\label{Amp-3-2}
{\cal M}_{\mu\nu} \Big|_{\mathbb{S}=\mathbb{I}}=
\int (d^4 k) \,\text{tr} \big[ E_{\mu\nu}(k) \, \gamma^- \big]
\underbrace{
\int (d^4 z) \, e^{ikz}
\langle P| \,   \textbf{:}  \bar\psi(0) \gamma^+ \psi(z)  \textbf{:} \,| P \rangle}_{\Phi^{[\gamma^+]}(k)}
\Big].
%\nonumber
\end{eqnarray}
Then,  we use the Fourier transforms for the operators in the correlator to derive the following
\begin{eqnarray}
\label{Amp-3-3}
\langle P| \,   \textbf{:}  \bar\psi(0) \gamma^+ \psi(z)  \textbf{:} \,| P \rangle
= \int (d^4 k_1 d^4 k_2) e^{ -i k_1 z}
\underbrace{ \big[ \bar u(k_2) \gamma^+ u(k_1) \big]}_{L^{[\gamma^+]}(k_2, k_1)} \,\,
\underbrace{\langle P | b^+(k_2) b^-(k_1) | P \rangle}_{\delta^{(4)}(k_1-k_2) {\cal M}(k_2,k_1 | P)},
%\nonumber
\end{eqnarray}
where $L^{[\gamma^+]}(k_2, k_1)$ gives the Lorentz structure (Lorentz parametrization) and
${\cal M}(k_2,k_1 | P)$ is the quark-hadron ${\cal M}$-amplitude.

After factorization at the leading order and in the collinear limit, we finally derive that
\begin{eqnarray}
\label{Amp-3-4}
{\cal M}_{\mu\nu} \Big|_{\mathbb{S}=\mathbb{I}}=
\int (dx) \,\text{tr} \big[ E_{\mu\nu}(xP^+) \, \gamma^- \big] \, \Phi^{[\gamma^+]}(x)
%\nonumber
\end{eqnarray}
where ($k=(k^+, k^-, \vec{\bf k}_\perp)$)
\begin{eqnarray}
\Phi^{[\gamma^+]}(x) = \int (d^4 k) \delta(x-k^+/P^+) \Phi^{[\gamma^+]}(k)
\stackrel{{\cal F}}{=}
\underbrace{
\langle P| \,   \textbf{:}  \bar\psi(0) \gamma^+ \psi(0^+,z^-, \vec{\bf 0}_\perp)  \textbf{:} \,| P \rangle.}
_{\sim\,\, \text{math. probability to find parton inside hadron}}
%\nonumber
\end{eqnarray}
\vspace{0.2cm}

Till now, we deal with the standard case of the collinear framework where the quark content of hadrons has been described
by the static probability amplitudes, {\it i.e.} the corresponding parton distributions stay without any evolutions.

%%%%%%%%%%%%%%%%%%%%%%%%%%%%%%%%%%%%
\section{Parton distributions from the correlators with interactions}
\label{PD-S}
%%%%%%%%%%%%%%%%%%%%%%%%%%%%%%%%%%%%

In this section, we discuss the effect of the interactions which has been accumulated in the
correlator. We distinguish two cases:
({\it i}) the interactions in the correlators have been included
provided rather a small transverse parton momenta, $|k_\perp|= {\cal O}(k^2_\perp)$;
({\it ii}) the interactions in the correlators have been taken into account with the essential $k_\perp$-dependence.
The loop integrations appeared at a given order of interactions give the explicit (or evolution)
and implicit (or structure) integrations, see Fig.~\ref{F-1}.
Unlike the case ({\it ii}), the implicit loop integration of the case ({\it i}) cannot produce the new type of parton distributions with
the essential $k_\perp$-dependence.

%%%%%%%%%%%%%%%%%%%%%%%%%%%%%%%%%%%%
\subsection{The almost collinear limit, $k_\perp\approx 0$, and $\mathbb{S}\not=\mathbb{I}$}
\label{AlmColF}
%%%%%%%%%%%%%%%%%%%%%%%%%%%%%%%%%%%%

\begin{figure}[tbp]
\centering % \begin{center}/\end{center} takes some additional vertical space
\includegraphics[width=.49\textwidth]{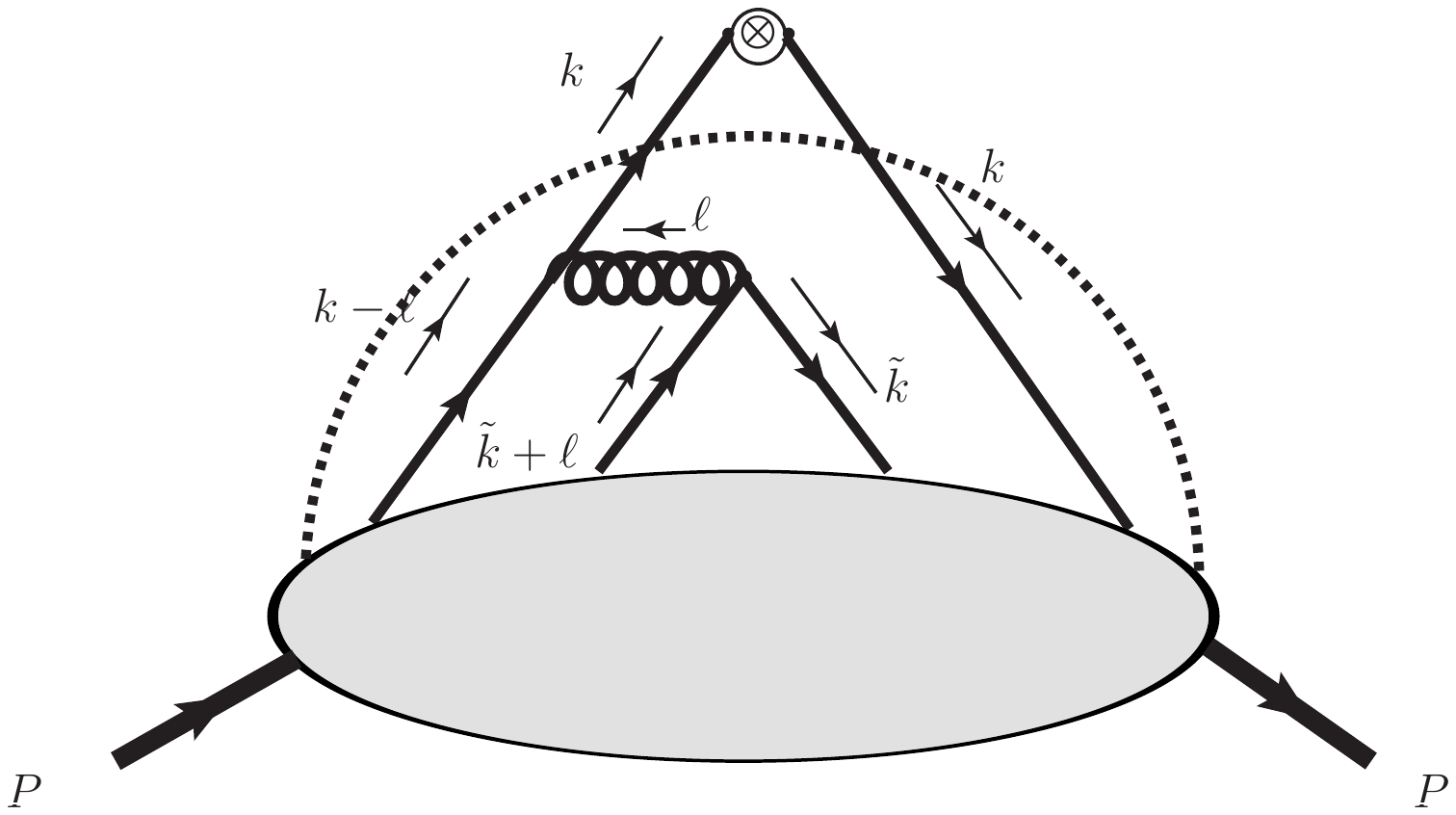}\,
\includegraphics[width=.49\textwidth]{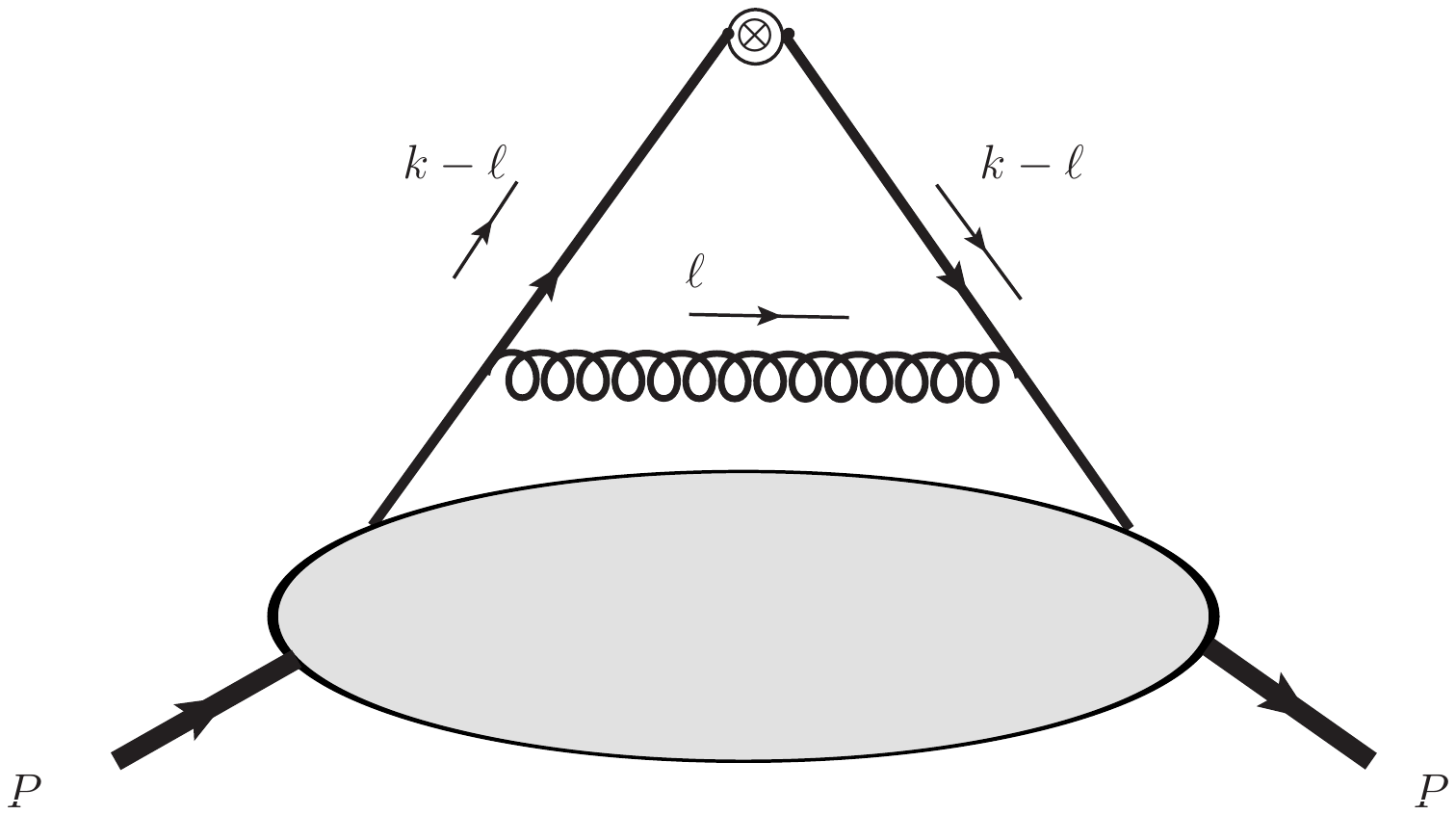}
% "\includegraphics" is very powerful; the graphicx package is already loaded
\vspace{-6cm}
\caption{The types of loop integrations in the corresponding correlators:
the left panel corresponds to the demonstration of the implicit loop integrations defined the Lorentz structure;
the right panel -- to the explicit lop integrations contributing to the evolution integration kernels.}
\label{F-1}
\end{figure}

First, we dwell on the conditions which provide the evolutions of well-known functions.
For this aim, we take into account the interaction but we still adhere the almost collinear case of
$k_\perp\approx 0$ ($k_\perp$-integrated functions):
\begin{eqnarray}
&&\Phi^{[\gamma^+]}(x) = \int (d^4 k) \delta(x-k^+/P^+) \Phi^{[\gamma^+]}(k)
\stackrel{{\cal F}}{=}
\nonumber\\
&&
\hspace{-3.5cm}
\xymatrix{
&& \boxed{
\langle P| \,\widetilde{{\rm T}} \, \bar\psi(0) \gamma^+ \psi(0^+, z^-, \vec{\bf 0}_\perp)
\mathbb{S}[\bar\psi, \psi, A] | P \rangle_c }
\ar@/_/@{=>}[dl] \ar@/^/@{=>}[dr]\\
& \hspace{1.5cm}\text{``Evolution''}  &&  \hspace{-0.8cm}\text{``Structure''}
}
%\nonumber
\end{eqnarray}
Here, ``Evolution'' implies the explicit loop integrations at the given order of interaction,
while ``Structure'' -- the implicit loop integration, see Fig.~\ref{F-1}.

Notice that in the considered case of a small $k_\perp$ the implicit loop integration at the given order of interactions,
which forms the spinor structure, is very limited and we do not expect any new parametrizing functions.
Moreover, we can observe that in this case the Lorentz
parametrization is, roughly speaking, ``commutative'' regarding the explicit (evolution) loop integration.
We call this case as the standard one.
We stress that, following the standard way, we can first consider the relevant operators as free operators in order to make the corresponding
parametrization. Then, $\mathbb{S}$-matrix can be included to derive the evolution
of the already-introduced parametrizing functions.

 %
%
%%%%%%%%%%%%%%%%%%%%%%%%%%%%%%% FIGURE %%%%%%%%%%%%%%%%%%%%%%%%%%%
%%%%%%%%%%%%%%%%%%%%%%%%%%%%%%%%%%%%%%%%%%%%%%%%%%%%%%%%
%%%%%%%%%

%%%%%%%%%%%%%%%%%%%%%%%%%%%%%%%%%%%%
\subsection{The case of  $k_\perp \not= 0$, and $\mathbb{S}\not=\mathbb{I}$}
\label{NotColF}
%%%%%%%%%%%%%%%%%%%%%%%%%%%%%%%%%%%%

However, if  the interaction encoded in the correlator
gives the essential $k_\perp$-dependence ($k_\perp$-unintegrated functions), we have
\begin{eqnarray}
&&\Phi^{[\gamma^+]}(x, k_\perp) = \int (dk^+ dk^-) \delta(x-k^+/P^+) \Phi^{[\gamma^+]}(k)
\stackrel{{\cal F}}{=}
\nonumber\\
&&
\hspace{-3.5cm}
\xymatrix{
&& \boxed{
\langle P| \,\widetilde{{\rm T}} \, \bar\psi(0) \gamma^+ \psi(0^+, z^-, \vec{\bf z}_\perp)
\mathbb{S}[\bar\psi, \psi, A] | P \rangle_c }
\ar@/_/@{=>}[dl] \ar@/^/@{=>}[dr]\\
& \hspace{1.5cm}\text{``Evolution''}  &&  \hspace{-0.8cm}\text{``Structure''}
}
%\nonumber
\end{eqnarray}
In contrast to the previous case, the explicit (evolution) loop integration and the Lorentz parametrization are now not
``commutative'' in a sense that we have first to implement the parametrization and, then, to study the corresponding evolutions.
In the $k_\perp$-dependent case, the implicit (structure) loop integration  gives more possibilities for the parametrization
which ultimately lead to the presence of new functions.
We call the present case as the non-standard one.
It is important to emphasize that in the non-standard way, we advocate in this paper,
we parametrize the given correlator where $\mathbb{S}$-matrix has been
presented from the very beginning.

%%%%%%%%%%%%%%%%%%%%%%%%%%%%%%%%%%%%
\section{Evidences for the new $k_\perp$-dependent functions}
\label{DerNF}
%%%%%%%%%%%%%%%%%%%%%%%%%%%%%%%%%%%%

In order to demonstrate the appearance of new functions, it is instructive
to begin with the well-know $k_\perp$-dependent function $f_1$ (see \cite{Anikin:2021zxl, Anikin:2022ocg, Anikin:2022eyf} for details):
\begin{eqnarray}
\label{Me-1}
\Phi^{[\gamma^+]}(k)=
P^+ f_{1} \left(x; \, k^2_\perp, (k_\perp P_\perp)\right)
\end{eqnarray}
where the parametrizing function $f_{1}$ depends on $k_\perp$ and the correlator corresponding to $\Phi^{[\gamma^+]}(k)$
includes the interactions.

Notice that the presence of transverse components in the parton momenta (or the deviation from the collinear motions of partons inside
hadrons) signals already on the influence of interactions. Indeed, let us decompose the $k_\perp$-dependent function $f_1$ over the
transverse components of momentum, we get that
\begin{eqnarray}
\label{Me-1-2}
&&\Phi^{[\gamma^+]}(k)=
\nonumber\\
&&
P^+ (k_\perp P_\perp) f_1^{(1)} (x; \,k^2_\perp) +
\Big\{ \text{terms of} \,\,(k_\perp P_\perp)^n  \,| \, n=0, n\ge 2  \Big\},
%\nonumber
\end{eqnarray}
where $k=(xP^+, k^-, \vec{\bf k}_\perp)$.
In Eqn.~(\ref{Me-1-2}), the function $f_{1} \left(x; \, k^2_\perp, (k_\perp P_\perp)\right)$
has been decomposed into the powers of $(k_\perp P_\perp)$.
Keeping the term of decomposition with $n=1$ represents the minimal necessary requirement for
the manifestation of new functions.

On the other hand, if we decompose $\mathbb{S}$-matrix and focus on the second order of strong interactions, $\mathbb{S}^{(2)}_{QCD}$, in the correlator,
we get the following
\begin{eqnarray}
\label{Me-1-3}
&&\langle P, S| T \bar\psi(0)\,\gamma^+\, \psi(z)\,\, \mathbb{S}_{QCD}[\bar\psi, \psi, A] |P, S \rangle =
\nonumber\\
&&
\langle P, S| T \bar\psi(0)\,\gamma^+\, \psi(z)\,\, \Big\{ \mathbb{I}  +
\mathbb{S}^{(2)}_{QCD}[\bar\psi, \psi, A]  +  .....\Big\}|P, S \rangle
\end{eqnarray}
where, as mentioned above, the term with $\mathbb{S}^{(2)}_{QCD}[\bar\psi, \psi, A]$
gives the evolution of previously-introduced parametrizing functions, see Fig.~\ref{F-1}, the right panel,
and the possibility to introduce the new function due to the implicit loop integrations which define the parton structure,
see Fig.~\ref{F-1}, the left panel.
Further, focusing on the contribution of diagram depicted in Fig.~\ref{F-1}, the left panel, we obtain
\begin{eqnarray}
\label{Me-2}
&&
\langle {\cal O}^{[\gamma^+]}\rangle^{(2)} \equiv
\langle P,S | T \bar\psi(0) \gamma^+ \psi(z) \, \mathbb{S}_{QCD}^{(2)}[\psi,\bar\psi, A] \,| P,S \rangle
\Big|_{\text{diagram of Fig.~{\ref{F-1}, the left panel}}} =
\nonumber\\
&&
\int (d^4 k) e^{-i (kz)} \Delta(k^2) \int (d^4 \ell) \Delta(\ell^2) \int (d^4 \tilde k)
 {\cal M}\left( k^2, \ell^2, \tilde k^2, ... \right)
\nonumber\\
&&\times
\big[ \bar u (k) \gamma^+ \hat k \gamma^\perp_\alpha u(k-\ell)\big]
\big[ \bar u (\tilde k) \gamma^\perp_\alpha u(\tilde k + \ell)\big],
%\nonumber
\end{eqnarray}
where
\begin{eqnarray}
\label{q-g-prop}
S(k)=\hat k \Delta(k^2), \,\,\hat k = (k\gamma), \quad D^\perp_{\mu\nu}(\ell)= g^\perp_{\mu\nu} \Delta(\ell^2),\,\,
\Delta(k^2) = \frac{1}{k^2 + i\epsilon}
%\nonumber
\end{eqnarray}
and ${\cal M}$-amplitude is given by
\begin{eqnarray}
\label{M-me}
{\cal M} \left( k_i^2,  (k_i k_j), ...\right) \delta^{(4)} (k_1+k_3 - k_2-k_4) =
\langle P,S | b^+(k_1) b^-(k_2) b^+(k_3) b^-(k_4) | P,S \rangle.
%\nonumber
\end{eqnarray}
We are now singling out the regions where $| \ell | \ll \{|k|, |\tilde k| \}$ and $|\tilde k| \sim |k|$.
After rather a simple spinor algebra which is based on the Fierz transformations denoted below as $Fi. \, tr.$
(we refer the reader to \cite{Anikin:2021zxl, Anikin:2022ocg, Anikin:2022eyf} for all details),
we can derive that
\begin{eqnarray}
\label{Me-2w}
&&
\langle P,S | T \bar\psi(0) \gamma^+ \psi(z) \, \mathbb{S}_{QCD}^{(2)}[\psi,\bar\psi, A] \,| P,S \rangle
\Big|_{\text{diagram of Fig.~{\ref{F-1}, the left panel}}}
\nonumber\\
&&
\stackrel{Fi. \, tr.}{
\Longrightarrow}
\big[ \bar u^{(\up_x)}(k) \gamma^+ \gamma^\perp \gamma_5 u^{(\up_x)}(k)\big]
\big[ \bar u^{(\up_x)}(k) u^{(\up_x)}(k)\big] \sim s_\perp.
\end{eqnarray}
We have shown that the $k_\perp$-dependence of $f_1$ has been originated from the intrinsic interactions encoded in the 
correlator.  Moreover, thanks to the mentioned interactions and the corresponding Fierz transforms 
the transverse momentum $k_\perp$ of Eqn.~(\ref{Me-1-2}) can be traced to  the quark transverse spin (axial)vector $s_\perp$,
see Eqn.~(\ref{Me-2w}).

As a result, the quark spin (axial)vector $s_\perp$ appears as the consequence of inner interactions encoded in the correlator and
$s_\perp$, as the corresponding inner parameter,
can participate in the
Lorentz parametrization, i.e.
\begin{eqnarray}
\label{Me-2w-2}
&&\langle P, S| {\cal O}^{[\gamma^+]} \big(0, z \,|\,  \bar\psi, \psi, A  \big) |P, S \rangle \Big |_{\mathbb{S}\not=\mathbb{I}}=
\nonumber\\
&&
\langle P, S|  T \bar\psi(0)\,\gamma^+\, \psi(z)\,\,
\mathbb{S}^{(2)}_{QCD}[\bar\psi, \psi, A] |P, S \rangle\Big|_{\text{diagram of Fig.~{\ref{F-1}, the left panel}}}
\stackrel{{\cal F}}{=}
i \epsilon^{+ - P_\perp s_\perp} \tilde f_1^{(1)} + ....
\end{eqnarray}

For the existence of Lorentz vector defined as $\epsilon^{+ - P_\perp s_\perp}$, it is necessary to assume that
the quark spin $s_\perp$ is not a collinear vector to the hadron transverse momentum $P_\perp$.
Within the Collins-Soper frame, the hadron transverse momentum can be naturally presented as
$\vec{\bf P}_\perp=(P_1^\perp, 0)$.
Since the hadron spin vector $S$ can be decomposed on the
longitudinal and transverse components as
$S^L + S^\perp = \lambda P^+/m_N + S^\perp$,
we get $P\cdot S = \vec{\bf P}^\perp\,\vec{\bf S}^\perp =0$. Hence, it is natural to suppose that
quark $s^\perp$ and hadron $S^\perp$ are collinear ones.
This is a kinematical constraint (or evidence) for the nonzero Lorentz combination
$\epsilon^{+ - P_\perp s_\perp}$
and, therefore, for the existence of a new function $\tilde f_1^{(1)} (x; \,k^2_\perp)$.

%%%%%%%%%%%%%%%%%%%%%%%%%%%%%%%%%%%%
\section{The principal result and applications}
\label{Res}
%%%%%%%%%%%%%%%%%%%%%%%%%%%%%%%%%%%%

So, it explicitly shows that the function $\tilde f_1^{(1)} (x; \,k^2_\perp)$ and its analogues must appear in the
parametrization of the hadron matrix element, {\it i.e.}
\begin{eqnarray}
\label{Phi-plus}
\hspace{-0.1cm}\Phi^{[\gamma^+]}(k)&\equiv&
\int (d^4 z) e^{+i(kz)} \langle P,S | \bar\psi(0) \, \gamma^+
\, \psi(z) \, \mathbb{S}[\psi,\bar\psi, A] \,
| P,S \rangle \Big|^{k^-=0, k_\perp\not=0}_{k^+=xP^+}
\nonumber\\
&=& i \epsilon^{+ - P_\perp s_\perp} {\tilde f}_1^{(1)} (x; \,k^2_\perp) +
i \epsilon^{+ - k_\perp s_\perp} f_{(2)}(x; \,k^2_\perp) + ....,
%\nonumber
\end{eqnarray}
where the ellipse denotes the other possible terms of parametrization.
We also observe that Lorentz structure tensor, $\epsilon^{+ - P_\perp s_\perp}$,
associated with our function
resembles the Sivers structure, $\epsilon^{+ - P_\perp S_\perp}$ in which
the nucleon spin vector $S_\perp$ is replaced by the quark spin vector $s_\perp$.

However, despite this similarity the Sivers function and the introduced function
${\tilde f}_1^{(1)} (x; \,k^2_\perp)$
have totally  different physical meaning.

The simplest example of application
is related to the well-known unpolarized Drell-Yan (DY) process, {\it i.e.}
the lepton-production in nucleon-nucleon collision:
\begin{eqnarray}
\label{DY-proc-st}
%&&
N(P_1) + N(P_2) \to \gamma^*(q) + X(P_X)
%\nonumber\\
%&&\hspace{2.65cm}
\to\ell(l_1)+\bar\ell(l_2) + X(P_X),
%\nonumber
\end{eqnarray}
with the initial unpolarized nucleons $N$.
The importance of the unpolarized DY differential cross section is due to
the fact that it has been involved in the denominators of any spin asymmetries.

At the leading order, the hadron tensor which
describes the unpolarized DY-process takes the following form:
\begin{eqnarray}
\label{h-t-4-1}
{\cal W}_{\mu\nu}^{(0)}= &&\delta^{(2)}(\vec{\bf q}_\perp)
\int (d x)  (d y) \delta(x P^{+}_1 - q^+)
\delta(yP^{-}_2 -q^-)
\nonumber\\
&&
\times
{\rm tr} \big[ \gamma_\nu \, \gamma^+ \,\gamma_\mu \, \gamma^- \big]\,
\Phi^{[\gamma^-]}(y) \Big\{ \int (d^2 \vec{\bf k}_1^\perp )\bar\Phi^{[\gamma^+]}(x, k^{\perp\,2}_1) \Big\},
%\nonumber
\end{eqnarray}
where
\begin{eqnarray}
\label{DY-funs}
\Phi^{[\gamma^-]}(y) = P_2^-\, f(y), \quad
\bar\Phi^{[\gamma^+]}(x, k^{\perp\,2}_1) =
i\epsilon^{+ - k_1^\perp s^\perp} f_{(2)} (x; \,k^{\perp\, 2}_1).
%\nonumber
\end{eqnarray}

Calculating the contraction of hadron tensor with the unpolarized lepton tensor ${\cal L}^U_{\mu\nu} $, we derive that
\begin{eqnarray}
\label{xsec-unpl}
&&d\sigma^{unpol.} \sim \int (d^2 \vec{\bf q}_\perp) {\cal L}^{U}_{\mu\nu}
{\cal W}^{(0)}_{\mu\nu} =
\\
&&
\int (d x)  (d y) \delta(x P^{+}_1 - q^+)
\delta(yP^{-}_2 -q^-)
%\nonumber\\
%&&
%\times
(1+\cos^2\theta) f(y)  \int (d^2 \vec{\bf k}_1^\perp ) \epsilon^{P_2 - k_1^\perp s^\perp} \Im{m} f_{(2)} (x; \,k^{\perp\, 2}_1),
\nonumber
\end{eqnarray}
where
\begin{eqnarray}
\epsilon^{+ - k_1^\perp s^\perp} = \vec{\bf k}_1^\perp \wedge \vec{\bf s}^{\perp} \sim \sin (\phi_k - \phi_s)
%\nonumber
\end{eqnarray}
with $\phi_A$, for $A=(k, s)$, denoting the angles between $\vec{\bf A}_\perp$ and $O\hat x$-axis
in the Collins-Soper frame.
Thus, the new $k_\perp$-dependent function $f_{(2)} (x; \,k^{\perp\, 2}_1)$ gives the additional and additive contribution to the
depolarization factor $D_{[1+\cos^2\theta]}$ appeared in the differential cross section of unpolarized DY process.

%%%%%%%%%%%%%%%%%%%%%%%%%%%%%%%%%%%%
\section{Conclusions}
\label{Conc}
%%%%%%%%%%%%%%%%%%%%%%%%%%%%%%%%%%%%

We have introduced the new $k_\perp$-dependent function ${\tilde f}_1^{(1)} (x; \,k^2_\perp)$
and $f_{(2)} (x; \,k_{\perp}^{2})$
which describe the transverse quark motion by the quark alignment along the fixed transverse direction.
The introduced functions can be considered as a ``in-between'' functions
of the Sivers and Boer-Mulders functions.
We have shown that, to the second order of strong interactions,
the new parametrizing function $\tilde f_1^{(1)} (x; \,k^2_\perp)$ can be related to
the function $f_1^{(1)} (x; \,k^2_\perp)$ imposing
the condition $\ell \ll |\tilde k| \sim |k|$ which corresponds to the regime where
the appeared two spinor lines are interacting by exchanging of soft gluon.
Moreover, the occurred four spinors generated by
two spinor lines have the polarizations aligned along the same
transverse direction.
In physical terms, the $k_\perp$-dependent function $\tilde f_1^{(1)} (x; \,k^2_\perp)$ which describes
the regime where
$k_\perp$-dependence (or the transverse motion of quarks inside the hadron) has been
entirely generated by the
quark spin alignment.

%%%%%%%%%%%%%%
\acknowledgments
%%%%%%%%%%%%%%

We thank the organizers of the “International Conference on Quantum Field Theory, High-Energy Physics, and Cosmology”
for the invitation to give a talk. Our special thanks go to
L.~Szymanowski for very fruitful and illuminating discussions.

%%%%%%%%%%%%%%%%%%%%%%


\begin{thebibliography}{99}
\vspace{1\baselineskip}

%\cite{Boglione:1999pz}
\bibitem{Boglione:1999pz}
M.~Boglione and P.~J.~Mulders,
%``Time reversal odd fragmentation and distribution functions in p p and e p single spin asymmetries,''
Phys. Rev. D \textbf{60}, 054007 (1999)

%\cite{Bacchetta:2004jz}
\bibitem{Bacchetta:2004jz}
A.~Bacchetta, U.~D'Alesio, M.~Diehl and C.~A.~Miller,
%``Single-spin asymmetries: The Trento conventions,''
Phys. Rev. D \textbf{70}, 117504 (2004)


%\cite{Goeke:2005hb}
\bibitem{Goeke:2005hb}
K.~Goeke, A.~Metz and M.~Schlegel,
%``Parameterization of the quark-quark correlator of a spin-1/2 hadron,''
Phys. Lett. B \textbf{618}, 90-96 (2005)

%\cite{Collins:2005rq}
\bibitem{Collins:2005rq}
J.~C.~Collins,A.~V.~Efremov, K.~Goeke, M.~Grosse Perdekamp, S.~Menzel, B.~Meredith, A.~Metz and P.~Schweitzer,
%``Sivers effect in Drell Yan at RHIC,''
Phys. Rev. D \textbf{73}, 094023 (2006)

%\cite{Anselmino:2008sga}
\bibitem{Anselmino:2008sga}
M.~Anselmino, M.~Boglione, U.~D'Alesio, A.~Kotzinian, S.~Melis, F.~Murgia, A.~Prokudin and C.~Turk,
%``Sivers Effect for Pion and Kaon Production in Semi-Inclusive Deep Inelastic Scattering,''
Eur. Phys. J. A \textbf{39}, 89-100 (2009)

%\cite{Bastami:2018xqd}
\bibitem{Bastami:2018xqd}
S.~Bastami, H.~Avakian, A.~V.~Efremov, A.~Kotzinian, B.~U.~Musch,
B.~Parsamyan, A.~Prokudin, M.~Schlegel, G.~Schnell and P.~Schweitzer, \textit{et al.}
%``Semi-Inclusive Deep Inelastic Scattering in Wandzura-Wilczek-type approximation,''
JHEP \textbf{06}, 007 (2019)

%\cite{Boer:1997nt}
\bibitem{Boer:1997nt}
D.~Boer and P.~J.~Mulders,
%``Time reversal odd distribution functions in leptoproduction,''
Phys. Rev. D \textbf{57}, 5780-5786 (1998)

%\cite{Efimov:1993zg}
\bibitem{Efimov:1993zg}
G.~V.~Efimov and M.~A.~Ivanov,
\emph{The Quark confinement model of hadrons},
\emph{Publisher: IOP}  (1993)


%\cite{Anikin:2021zxl, Anikin:2022ocg, Anikin:2022eyf}
%\cite{Anikin:2021zxl}
\bibitem{Anikin:2021zxl}
I.~V.~Anikin and L.~Szymanowski,
%``Alignment function as a new kind of transverse momentum dependent functions,''
Eur. Phys. J. A \textbf{58}, 160 (2022)

%\cite{Anikin:2022ocg}
\bibitem{Anikin:2022ocg}
I.~V.~Anikin and L.~Szymanowski,
\emph{``Archetypal Factorization and Gluon Poles in semi-exclusive reactions,''}
[arXiv:2202.09218 [hep-ph]].


%\cite{Anikin:2022eyf}
\bibitem{Anikin:2022eyf}
I.~V.~Anikin and L.~Szymanowski,
%Parton distributions: Functional complexity and Lorentz parametrization},
 Inter. Journal of Modern Physics A, 2250139 (2022)
[arXiv:2204.07510 [hep-ph]]

\end{thebibliography}
\end{document}